\documentclass[
    10pt,
    floatfix,
    twocolumn,
    superscriptaddress,
    groupedaddress,
    prb,
    a4paper,
    showpacs
]{revtex4}

\usepackage{graphics,graphicx}
\usepackage{amssymb,amsmath}
\usepackage{color}
\usepackage{bm}
\usepackage{float}
\usepackage{epsfig}
\usepackage[T1]{fontenc}
\newcommand{\w}{\omega}
\newcommand{\igr}{\includegraphics}
\newcommand{\df}{\partial}

\newcommand{\real}{\mathrm{Re}}
\newcommand{\imag}{\mathrm{Im}}
\newcommand{\etal}{\mbox{\textit{et al.}}}

\newcommand{\correlator}[1]{\langle#1\rangle}

\newcommand{\beq}{\begin{equation}}
\newcommand{\eeq}{\end{equation}}
\newcommand{\beqa}{\begin{eqnarray}}
\newcommand{\eeqa}{\end{eqnarray}}

\newcommand{\mev}{\textrm{meV}}
\newcommand{\muev}{\textrm{$\mu$eV}}

\newcommand{\ev}{\textrm{eV}}

\newcommand{\Ang}{\textrm{\AA}}
\newcommand{\Ry}{\textrm{Ry}}
\newcommand{\ps}{\textrm{ps}}

\newcommand{\W}{\textbf{K}}

\newcommand{\D}{\textbf{D}}
\newcommand{\I}{\textbf{I}}
\newcommand{\M}{\textbf{M}}
\newcommand{\X}{\textbf{X}}
\newcommand{\PI}{\bm{\Pi}}
\newcommand{\del}{\gamma_\lambda}

\newcommand{\equref}[1]{Eq.~(\ref{#1})}
\newcommand{\figref}[1]{Fig.~\ref{#1}}
\newcommand{\tabref}[1]{Table~\ref{#1}}
\newcommand{\secref}[1]{Sec.~\ref{#1}}
\newcommand{\appref}[1]{App.~\ref{#1}}

\newcommand{\Figref}[1]{Figure~\ref{#1}}
\newcommand{\Tabref}[1]{Table~\ref{#1}}

\begin{document}
\title{Atomistic theory for the damping of vibrational modes in mono-atomic gold chains}



\author{M. Engelund} \affiliation{DTU Nanotech,  Department of
Micro and Nanotechnology, Technical University of Denmark,
{\O}rsteds Plads, Bldg.~345E, DK-2800 Kongens Lyngby, Denmark}
\email{mads.engelund@nanotech.dtu.dk}

\author{M. Brandbyge}
\affiliation{DTU Nanotech, Department of Micro and Nanotechnology,
Technical University of Denmark, {\O}rsteds Plads, Bldg.~345E,
DK-2800 Kongens Lyngby, Denmark}

\author{A.~P. Jauho}
\affiliation{DTU Nanotech, Department of Micro and Nanotechnology,
Technical University of Denmark, {\O}rsteds Plads, Bldg.~345E,
DK-2800 Kongens Lyngby, Denmark} \affiliation{Department of Applied
Physics, Helsinki University of Technology, P.O.Box 1100, FI-02015
TKK, Finland}

\date{\today}
\begin{abstract}

We develop a computational method for evaluating the damping of vibrational modes in
mono-atomic metallic chains suspended between bulk crystals under external strain. The damping is due to the coupling
between the chain and contact modes and the phonons in the bulk substrates. The geometry
of the atoms forming the contact is taken into account. The dynamical matrix is computed
with density functional theory in the atomic chain and the contacts using finite atomic
displacements, while an empirical method is employed for the bulk substrate. As a
specific example, we present results for the experimentally realized case of gold chains
in two different crystallographic directions. The range of the computed damping rates
confirm the estimates obtained by fits to experimental data [Frederiksen~\etal, Phys.
Rev. B \textbf{75}, 205413(R)(2007)]. Our method indicates that an order-of-magnitude
variation in the harmonic damping is possible even for relatively small changes in the strain.  Such detailed
insight is necessary for a quantitative analysis of damping in metallic atomic chains,
and in explaining the rich phenomenology seen in the experiments.
\end{abstract}

\maketitle

\section{Introduction}
\label{sec:introduction}

The continuing shrinking of electronic devices and the concomitant great interest in
molecular electronics\cite{Cuniberti2005} have underlined the urgency of a detailed
understanding of transport of electrons through molecular-scale contacts. A particularly
important issue concerns the energy exchange between the charge carriers and the
molecular contact. Thus, the local Joule heating resulting from the current passing
through the contact, and its implications to the structural stability of such contacts
are presently under intense
investigation\cite{ScFrGa.2008,TeHoHa.2008,HuCh.2007,Galperin2007a,Ryndyk2008}.
Experimentally, local heating in molecular conductors in the presence of the current has
been inferred using two-level fluctuations\cite{TsTaKa.2008} and Raman
spectroscopy\cite{Ioffe2008}.

Mono-atomic chains of metal
atoms\cite{Bollinger2001} are among the simplest
possible atomic-scale conductors. The atomic gold chain is probably
the best studied atomic-sized conductor, and a great deal of
detailed information is available from
experiments\cite{Rodrigues2001a,Agrait2002,Agrait2002a,Legoas2002,Agrait2003,
Rego2003,Coura2004,Bettini2005,
Lagos2007,Hasmy2008,Kizuka2008,Thiess2008,Tsutsui2008a}, and related
theoretical studies\cite{Todorov1998,Bahn2001,Silva2001,Legoas2002,Rego2003,Chen2003,Montgomery2003,Frederiksen2004,Viljas2005,
Paulsson2005,Frederiksen2007a,
Frederiksen2007,Lagos2007,Hobi2008}. The current
induced vibrational excitation and the stability of atomic metallic
chains have been addressed in a few
experiments\cite{Yasuda1997,SmUnRu.2004,Tsutsui2005,TsKuSa.2006}.

In the case of a gold chain Agra{\"i}t \etal\cite{Agrait2002a} reported well-defined
inelastic signals in the current-voltage characteristics. These signals were seen as a
sharp 1\% drop of the conductance at the on-set of back-scattering due to vibrational
excitation when the voltage equals the vibrational energy. Especially for the longer
chains (6-7 atoms), the vibrational signal due to the Alternating Bond-Length (ABL)
mode\cite{Frederiksen2004,Frederiksen2007a}, dominates. This resembles the situation of
an infinite chain with a half-filled electronic band where only the zone-boundary phonon
can back-scatter electrons\cite{Agrait2002} due to momentum conservation.

The inelastic signal gives a direct insight into how the frequency
of the ABL-mode depends on the strain of the atomic chain. This
frequency can also be used to infer the bond strength. The signature
of heating of the vibrational mode is the non-zero slope of the
conductance versus voltage beyond the on-set of excitation: with no
heating the curve would be flat.  Fits to the experiment on gold
chains using a simple model\cite{Paulsson2005} suggest that the
damping of the excitation, as expected, can be significant. However,
the experiments in general show a variety of behaviors and it is not
easy to infer the extent of localization of the ABL vibration or its
damping in these systems \footnote{N. Agra\"it, private
communications.}.

In order to address the steady-state effective
temperature of the biased atomic gold chain theoretically, it is necessary to
consider the various damping mechanisms affecting the localized
vibrations, such as their coupling to the vibrations in the contact,
or to the phonons in the surrounding bulk reservoirs. This is the
purpose of the present paper: we calculate the vibrational modes in
atomic gold chains and their coupling and the resulting damping due
to the phonon system in the leads.  We work within the harmonic
approximation and employ first principles density functional theory
(DFT) for the atomic chain and the
contacts\cite{SoArGa.02.SIESTAmethodab} while a potential model is
used for the force constants of the leads\cite{Treglia1985}.

Experimental TEM studies\cite{Rodrigues2001,Kizuka2008} have shown
that atomic chains form in the $\langle 100 \rangle$ and $\langle
111 \rangle$ directions while the $\langle 110 \rangle$ direction
gives rise to thicker rods\cite{Rodrigues2001}. Therefore we focus
on chains between two (100)-surfaces or (111)-surfaces. We consider
chain-lengths of 3-7 atoms and study the behavior of their
vibrations and damping when the chains are stretched. The TEM
micrographs also indicate that the chains are suspended between
pyramids, so in our calculations we add the smallest possible
fcc-stacked pyramid to link the chain to the given surfaces.

As we shall show below,  at {\it low strain} the gold chains have
harmonically undamped ABL-modes with frequencies outside the bulk
band. The long chains of 6-7 atoms also have ABL-modes with very low
damping at {\it high strain }. Our results indicate that chains between
(111)-surfaces will have a lower damping than chains between
(100)-surfaces. Importantly, we find that the damping is an
extremely sensitive function of the external strain: an order of
magnitude change may result from minute changes in the strain.  This
may provide a key for understanding the rich behavior found in
experiments.

The paper is organized as follows. In Sect. II we describe  how the
central quantities, i.e., the dynamical matrix,  the projected
density-of-states, and the damping rates are calculated. Section III
is devoted to the analysis of the numerical results we have
obtained, beginning with results for the structure of the chains,
proceeding to the dynamical matrix, and concluding with an analysis
of the damping of modes in the systems. Section IV gives our final
conclusions, while certain technical details are presented in three
appendices.

\section{Method}
\label{sec:method}
\begin{figure}[t]
  \includegraphics[width=.3\textwidth]{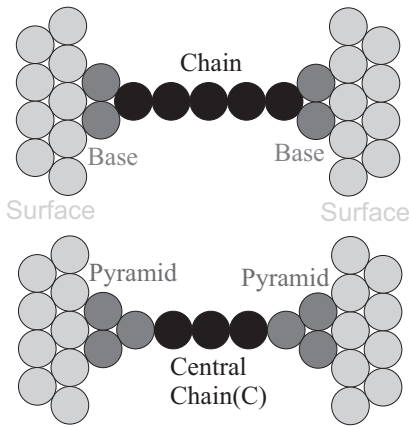}
  \caption{Two ways of partitioning the central part of the chain-substrate system.
  [Top]
  The Chain is the part of the system that only contains one atom in a plane parallel
  to the surface, and the Base is what connects the two-dimensional surface to the Chain. [Bottom]
  The Pyramid is the Base plus the Chain atom closest to the Base\protect.
  The Central Chain is the remaining part of the Chain after removing one atom at each end.
}
  \label{fig:partitionings}
\end{figure}

As will become evident in the forthcoming discussion it is
advantageous to use two different ways to label the atoms forming
the junction; these two schemes are illustrated in Fig.
\ref{fig:partitionings}.   The first scheme (Fig.
\ref{fig:partitionings} top panel) is based on the cross-sectional area and collects all atoms with
equilibrium positions on the one-dimensional line joining the two
surfaces into a "Chain", and calls the remaining atoms between the
Chain and the substrate the "Base". The
second scheme (Fig. \ref{fig:partitionings}, bottom) distinguishes
between a "Pyramid" and a "Central Chain"; this is chosen because
the last atom of the Chain has bonds to four or five atoms making this atom
very different from the Central Chain atoms that only have two bonds
per atom.

A quantity of central importance to all our analysis is the
mass-scaled dynamical matrix, $\W$, which we here define as including $\hbar$,
\begin{eqnarray}
  \W_{ij}=\frac{\hbar^2}{\sqrt{m_im_j}}\frac{\df^2 E}{\df u_i\df u_j}\quad ,
\end{eqnarray}
where $E$ is the total energy of the system, $u_i$ is the coordinate
corresponding to the $i$'th degree of translational freedom for the
atoms of the system. $m_i$ is the mass of the atom that the $i$'th degree freedom belongs to. $\W$ governs the the evolution of the vibrational system within the harmonic approximation. In the Fourier domain the Newton equation of motion reads
\begin{equation}
  \label{eq:1}
    \W u_\lambda=\epsilon^2_\lambda u_\lambda\quad ,
\end{equation}
where $\lambda$ denotes a mode of oscillation in the system and
$\epsilon_\lambda$ is the corresponding quantization energy.

\begin{figure}[t]
  \centering
  \includegraphics[width=.3\textwidth]{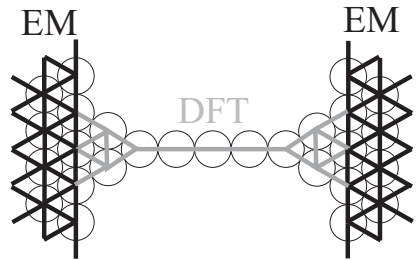}
  \caption{Parameters used for calculating the dynamical matrix.
  Only nearest-neighbor coupling is shown. The coupling elements labeled 'EM' are
  found by the empirical model and the coupling elements labeled 'DFT' by DFT.
  On-site elements are determined from the coupling elements (see \secref{sec:constr-dynam-matr})
  and are not shown in the figure.}
  \label{fig:partitionings3}
\end{figure}

 The evaluation of $\W$ proceeds as follows. Finite difference DFT
calculations (for details of our implementation, see
\appref{sec:constr-dynam-matr}) were used for the Chain, the Base
and the coupling between the surface and the Base while for the
surfaces we used an empirical model due to Tréglia and
Desjonquères\cite{Treglia1985}. \Figref{fig:partitionings3}
illustrates the domains for the two different methods. The position
of the interface between the region treated by DFT and the region
treated by the empirical model is a parameter that can be varied,
and the dependence on the final results of the choice of this
parameter is analyzed in \appref{sec:test-convergence}.

The empirical model can be used to describe the on-site and coupling
elements of atoms in a crystal structure. The model uses the bulk
modulus of gold to fit the variation in the force constant with
distance between nearest and next-nearest neighbors. Even though the
empirical model is fitted to the bulk modulus, which is a low
frequency property, it still accurately predicts the cutoff of the
bulk band. The positions of the neighboring atoms can only have
small deviations from perfect crystal positions (eg. bulk, surface
and ad-atoms). Note that this model is general enough to give
different coupling elements between surface atoms and bulk atoms.
The model also distinguishes between the coupling between surface
atoms with or without extra atoms added to the surface.

The DFT calculations were done with the SIESTA code, using the
Per\-dew-Burke-Enzerhof version of the GGA
ex\-change-corre\-la\-tion potential with stan\-dard norm-conserving
Troul\-lier-Martins pseuodopotentials. We used an SZP basis set with
a confining energy of $0.01~\Ry$. A mesh cutoff of $150~\Ry$ was
used. Relaxation was done with a force tolerance of
$0.002~\ev/\Ang$. These values were found to have converged for the
same type of system by Frederiksen
\etal\cite{Frederiksen2007a,Frederiksen2007}. The experimental fcc
bulk lattice constant of $4.08~\Ang$ was used. Only the device
region was relaxed (defined in \figref{fig:partitionings2}).

For the (111) orientations a 4$\times$4 atom surface unit and a
2$\times$3 $k$-point sampling was used, while for the (100)
orientations  a 3$\times$3 atom surface unit cell and a 3$\times$3
$k$-point sampling was employed. This ensured a similar and
sufficient $k$-point density for both kinds of surfaces (see \appref{sec:test-convergence}).

\subsection{Green's function for a perturbation on the surface}

\begin{figure}[ht]
  \centering
  \includegraphics[width=.3\textwidth]{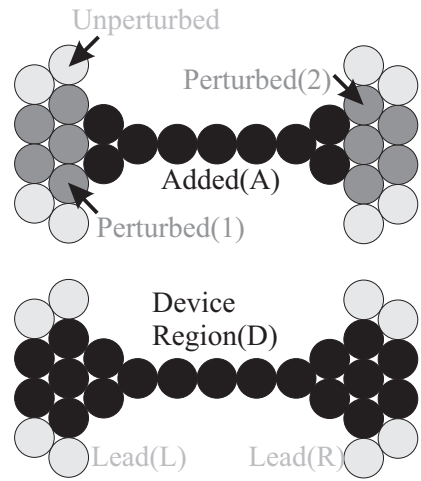}
  \caption{Adding atoms to two surfaces. [Top] The forces between surface atoms
  within next-nearest neighbor distance(4.08\Ang) of the added atoms are
  perturbed by the presence of the added atoms. [Bottom] The device region
  is where the coupling between the atoms is different from the values for
  the two unperturbed surfaces. The coupling between the device region and
  the leads is considered to be unperturbed.}
  \label{fig:partitionings2}
\end{figure}

All properties of interest in the present context can be derived
from the (retarded) Green's function  $\D$, defined by
\begin{equation}
  \label{eq:broad2}
  [(\epsilon+i\eta)^2\I-\W]\D(\epsilon)=\I\equiv \M \D\quad,
\end{equation}
where  $\eta=0^+$ and we defined the inverse of the Green's function
by $\M=\D^{-1}$. Specifically, we shall need the Green's function
projected onto the region close to the Chain. Our procedure is based
on a method due to  Mingo \etal\cite{Mingo2008} which has previously been tested in an investigation of finite Si nanowires between Si
surfaces. We define $\textbf{X}_{YZ}$ as the the
block of the matrix $\textbf{X}$, where the indices run over the
degrees of freedom in regions $Y,Z$, respectively, where $Y,Z =
\{1,2,A,D,L,R\}$, as defined either in
\figref{fig:partitionings} or \figref{fig:partitionings2}.

First, let us start with two perfect surfaces. We then add the atoms
that connect these surfaces (the Base and the Chain). Within a
certain range from the added atoms the on-site and coupling elements
of $\W$ will be different from the values for the perfect surface.
Together, the added atoms and the perturbed atoms define the device
region $D$ (\figref{fig:partitionings2}, bottom).
The coupling between the device region and the rest of the surface
($L,R$ for the left and right leads, respectively) is assumed to be
unperturbed.

In order to compute the Green's function projected on the device
region, $\D_{DD}(\epsilon)$, we first consider this matrix
representation of \equref{eq:broad2}\footnote{Formally, this equation
is derived by inserting identity operators $I\equiv |D\rangle\langle
D|+|\alpha\rangle\langle\alpha|$ in Eq.(\ref{eq:broad2}), and using
the basis $\{|D\rangle,|\alpha\rangle\}$ for the matrix
representation.}:
\begin{equation}
\left(
  \begin{array}{cc}
    \M_{DD} & \M_{D\alpha} \\
    \M_{\alpha D} & \M_{\alpha\alpha} \\
  \end{array}
\right) \left(
  \begin{array}{cc}
    \D_{DD} & \D_{D\alpha} \\
    \D_{\alpha D} & \D_{\alpha\alpha} \\
  \end{array}
\right)= \left(
  \begin{array}{cc}
    \I_{DD} & \textbf{0}_{D\alpha} \\
    \textbf{0}_{\alpha D} & \I_{\alpha\alpha} \\
  \end{array}
\right).
\end{equation}
Here the index $\alpha=(L,R)$, i.e., the left and right unperturbed
surface, while $D=\{1,A,2\}$. Using straightforward matrix
manipulations one finds
\begin{eqnarray}
  \label{eq:xx}
  \D_{DD}&=&[\M_{DD}-\M_{D\alpha}(\M_{\alpha\alpha})^{-1}\M_{\alpha D}]^{-1}\nonumber\\
         &=&[\M_{DD}-\PI_{DD}]^{-1}\quad,
\end{eqnarray}
which defines the self-energy
$\PI_{DD}=\M_{D\alpha}(\M_{\alpha\alpha})^{-1}\M_{\alpha D}$. Since
the added atoms do not couple to the unperturbed surfaces, and the
perturbed region 1 couples only to the right unperturbed surface
while the perturbed region 2 only couples to the left unperturbed
surface, the self-energy $\PI_{DD}$ has the matrix structure
\begin{equation}\label{eq:Pimatrix}
\PI_{DD}=\left(
      \begin{array}{ccc}
        \M_{1L}(\M_{LL})^{-1}\M_{L1} & 0 & 0 \\
        0 & 0 & 0 \\
        0 & 0 & \M_{2R}(\M_{RR})^{-1}\M_{R2} \\
      \end{array}
    \right).
\end{equation}
This object can be evaluated as follows. First, in the limit of
large regions 1 and 2, the coupling elements $\M_{L1}$ and $\M_{R2}$
must approach those of the unperturbed surface, $\M^S_{L1}$ and
$\M^S_{R2}$, respectively. In what follows, we shall make the
approximation that the regions 1 and 2 are chosen so, that this
condition is satisfied sufficiently accurately. Second, we note that
the matrix $\M_{\alpha\alpha}$ is {\it indistinguishable} from the
matrix $\M^S_{\alpha\alpha}$, as long as the involved atoms are
outside the perturbed regions 1 or 2. Therefore, we can write
\begin{eqnarray}\label{eq:selfenergy1}
\M_{1L}(\M_{LL})^{-1}\M_{L1}&\simeq&
\M^S_{1L}({\M^S_{LL}})^{-1}\M^S_{L1}\equiv\PI^S_{11}\nonumber\\
\M_{2R}(\M_{RR})^{-1}\M_{R2}&\simeq&
\M^S_{2R}({\M^S_{RR}})^{-1}\M^S_{R2}\equiv\PI^S_{22},\nonumber\\
\end{eqnarray}
where the accuracy increases with increasing size of regions 1 and
2.  On the other hand, using the definition of the self-energy, we
can write
\begin{eqnarray}\label{eq:selfenergy2}
\PI^S_{11}&=&\M^S_{11}-(\D^S_{11})^{-1}\nonumber\\
\PI^S_{22}&=&\M^S_{22}-(\D^S_{22})^{-1},
\end{eqnarray}
where $\D^S_{ii},i=1,2$ is the projection of the {\it unperturbed}
Green's functions onto the atoms in regions 1,2, respectively. This
object is evaluated by exploiting the periodicity in the ideal
surface plane. The Fourier transform of $\M^S$ in the parallel directions has
a tridiagonal block structure and we can solve for its inverse very
effectively using recursive techniques (see e.g. Sancho
\etal\cite{Sancho1984}). Of course we still have to evaluate the
Fourier transform for a large number of $k$-points. The density of
$k$-points as well as the size of the infinitesimal $\eta$ are
convergence parameters which determine the accuracy and cost of the
computation.  An analysis of the choice of these parameters is given
in \appref{sec:test-convergence}.

To sum up, the calculation is preformed in the following steps: (i)
Start with perfect leads and specify the device in between them.
(ii) The atoms in the leads where $\W$ is perturbed by the presence
of the device are identified. (iii) The unperturbed surface Green's
function $\D^S$ is found via $k$-point sampling and then used to
construct the self-energy,
Eqs.(\ref{eq:selfenergy1}--\ref{eq:selfenergy2}). (iv) The perturbed
Green's function is then found using this self-energy via
Eqs.(\ref{eq:xx}--\ref{eq:Pimatrix}).

\subsection{Modes and life-times}
\label{sec:life-times} For any finite system the eigenvalues
$\epsilon_\lambda^2$  and thereby also the density of states  are
found straightforwardly. For infinite systems we use that each
eigenvector, $u_\lambda$, with the corresponding eigenvalue,
$\epsilon_\lambda$ gives a contribution to the imaginary part of the
Green's function in the $\epsilon_\lambda\gg \eta$ limit

\begin{equation*}
  \label{eq:9}
u_\lambda^\dagger\imag\D(\epsilon)u_\lambda\approx-\frac{1}{2\epsilon_\lambda}\frac{1}{(\epsilon-\epsilon_{\lambda})^2+\eta^2}.
\end{equation*}

This expression  results in the following density of states
\begin{equation}
  \label{eq:4}
  \textbf{n}(\epsilon)=-\frac{2\epsilon}{\pi}\lim_{\eta\rightarrow
  0^+}\imag\D(\epsilon).
\end{equation}

The broadened vibrational modes of the device region can each be associated with a finite life-time. To do this we need to have a definition of an approximate
vibrational mode of the central part of the system that evolves into
an  eigenmode of $\W$  when the coupling to the leads tends to zero. We
define 'modes' as the vectors that for some energy, $\epsilon^*$,
correspond to a zero eigenvalue mode of $\real \D_{DD}(\epsilon^*)$
(see \secref{sec:finding-modes-peak} for details).

We also need to define a few characteristics of a mode. The Green's function projected
onto a mode can be approximated by a broadened free phonon propagator with
constants $\epsilon_\lambda$ and $\del$ in a neighborhood of the mode peak energy,
\begin{eqnarray*}
u_\lambda^\dagger\D_{DD}(\epsilon)u_\lambda&=&\frac{1}{(\epsilon+i\del)^2-\epsilon_\lambda^2}\\
&=&\frac{1}{\epsilon^2-(\epsilon_\lambda^2+\del^{2})+i2\epsilon \del}\quad.\\
\end{eqnarray*}
The time-dependent version of the Green's function is an
exponentially damped sinusoidal oscillation with damping rate of
$\frac{\del}{\hbar}$, mean life-time,
$\tau_\lambda=\frac{\hbar}{\del}$, and $Q$-factor,
$Q_\lambda=\frac{\epsilon_\lambda}{2\del}$. Comparing the broadened
phonon propagator to \equref{eq:xx} we see that
$u_\lambda^\dagger\imag\PI(\epsilon)u_\lambda=-2\epsilon\del$,
leading to
\begin{equation*}
  \label{eq:lifetime}
  \del=-\frac{u_\lambda^\dagger\imag\PI(\epsilon^*)u_\lambda}{2\epsilon^*}\quad,
\end{equation*}
where $\epsilon^*$ is the the mode peak energy.

This calculation of $\del$, $Q_\lambda$ and $\tau_\lambda$ only
strictly makes sense for peaks with a Lorentzian line shape. This
requires that $u_\lambda^\dagger\imag\D_{DD}(\epsilon)u_\lambda$ is
approximately constant across the peak which is the case for modes
with small broadening and large life-time. Nevertheless, we will
also use these definitions for the delocalized modes since the
calculated values are still a measure of interaction with the leads.

We also define a measure of spatial localization, $s_\lambda$,
\begin{equation*}
  \label{eq:2}
  s_\lambda=\frac{\sum_{x\in C} |(u_\lambda)_x|^2}{\sum_{x\in D\backslash C} |(u_\lambda)_x|^2}\frac{N_D-N_C}{N_C}\quad,
\end{equation*}
where $N_D$ and $N_C$ are the number of atoms in the device and central chain region
respectively and $D\backslash C$ means Device region except the Central Chain (the
perturbed reigion). This quantity is useful to pick out modes with a large amplitude in
the Central Chain region only. We have that $s_\lambda=1$ signifies equal amplitude in
$C$ and connecting atoms, while the limit $s_\lambda\rightarrow\infty$
($s_\lambda\rightarrow 0$) signifies a mode which is completely residing inside(outside)
the Chain.

It should be stressed that the mode properties calculated in this way only refer to
the harmonic damping by the leads and that other sources of damping are not included
such as electron-hole pair creation and anharmonicity. The damping due to electron-hole
pair creation, obtained by an {\it ab-initio} calculation on a selection of gold chains,
is about $50-80~\muev$ for the vibrational mode with the strongest coupling to
electrons\cite{Paulsson2005,Frederiksen2007a}. This type of damping is less dependent on
strain in gold chains due to stable electronic structure as evidenced by the robust
electronic conductance of one conductance quantum. The harmonic damping due to the leads
is typically higher than this, but as we shall see it can actually drop well below this
value and thus be less than the electron-hole pair damping.

In case of an applied bias the high frequency modes may be excited to a high occupation.
The creation of vibrational quanta is roughly proportional to $eV-\hbar\w_{\lambda}$,
while the damping mechanisms are not expected to have a strong dependence of the bias.
Therefore, as the bias is increased beyond the phonon energy threshold, the mode
occupation will rise and anharmonic interactions may become increasingly important even
for low temperatures. Mingo\cite{Mingo2006} has studied anharmonic effects on
heat-conduction in a model atomic contact, and more recently Wang and
co-workers\etal\cite{Wang2008} has used ab-initio calculations to access the effect of
anhamonicity on heat-conduction in carbon-based systems. Anharmonic effects are,
however, outside the scope of the present work.

\section{Results}
\label{sec:results}

\subsection{Geometrical Structure and the Dynamical Matrix}
\label{sec:life-times-primary}

In this subsection we investigate the geometrical structure of the
chains and the behavior of the dynamical
matrix. For each type of calculation (identified by the number of atoms in the Chain, the surface orientation and the type of Base) a range of calculations were set
up with the two surfaces at different separations, $L_i\,(i=1,2...)$,
with the separations incremented in equally spaced steps.
Trial-and-error was used to determine suitable step sizes for the
different types of calculations.

\begin{figure}
  \centering
  \igr[width=.45\textwidth,clip]{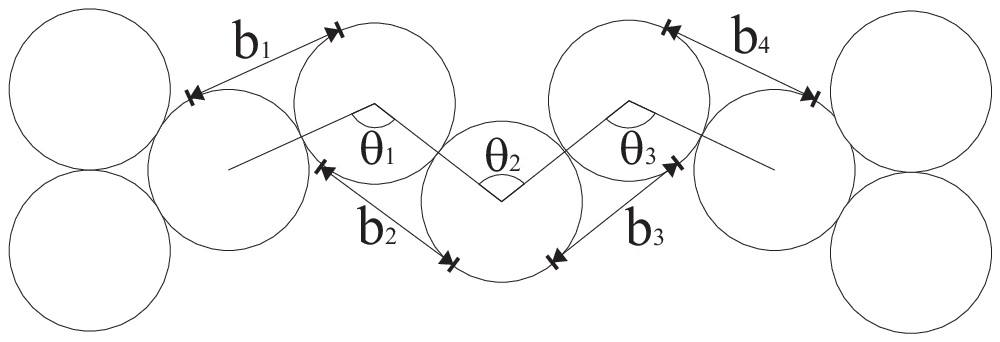}
  \caption{Distances and angles used to define the average bond length, $B=\correlator{b_j}$, and the average bond angle, $T=\correlator{\theta_j}$, respectively}
  \label{fig:bond_length_angle}
\end{figure}

To be able to compare chains of different lengths and between
different surfaces we define the average bond length,
$B=\correlator{b_j}$, as the average length between neighboring
atoms within the Chain, where $j$ runs over the number of bonds in
the Chain (see \figref{fig:bond_length_angle}). $B$ is useful
because it is closely related to the experimentally measurable
force\cite{Agrait2003} on the Chain and can be found without
interpolation. The close relationship between $B$ and the force is
demonstrated in \figref{fig:forces} where the force is calculated as
the slope of a least-squares fit of
\begin{equation*}
[(E_{i-1},L_{i-1}),(E_{i},L_{i}),(E_{i+1},L_{i+1})]
\end{equation*}
where $E$ is the total energy. We note that the force vs. $B$ curves to a good
approximation follows a straight line with a slope of $k=2.5~\ev/\Ang^2$, which can be
interpreted as the spring constant of the bonds in the Chain. In addition to $B$ we also
define the average bond angle $T=\correlator{\theta_j}$.

\begin{figure}
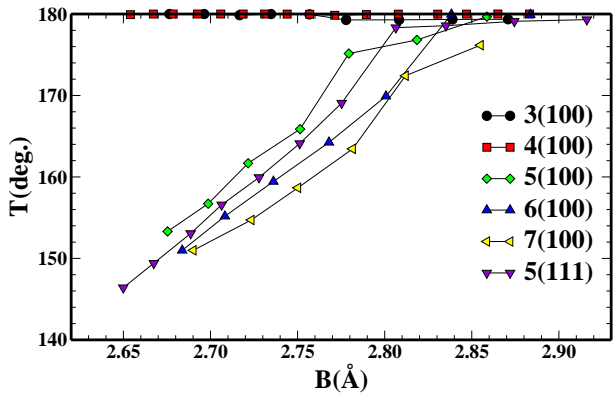

  \centering
  \igr[width=.45\textwidth,clip]{fig6}
  \caption{(color online) Average bond angle as a function of the average bond length.
  The long chains adopt a zig-zag structure at low $B$ while the short chains remain linear.}
  \label{fig:bond_angle}
\end{figure}

The behavior of the systems with respect to $B$ is relatively
simple. As the systems are strained it is mostly the bonds in the chain that are enlongated. Finally the central bond(s) become so
weak that they break. At low $B$ we see from \figref{fig:bond_angle}
that the longer chains adopt a zig-zag confirmation at low average
bond length. The 3- and 4-atom chains, however, remain linear within
the investigated range. Furthermore, the longer chains have a
similar variation in the average bond angle.

These preliminary observations are in agreement with previous theoretical studies by Frederiksen~\etal\cite{Frederiksen2007a} and S\'anchez-Portal~\etal\cite{Portal1999}. We recount these observations because we find that using $B$ as a parameter provides a helpful way to compare chain of different lengthts and because the calculations in this paper are the most accurate to date\footnote{The $k$-point sampling of Ref. [\onlinecite{Frederiksen2007a}] is so sparse that it may in certain instances give unrealistic predictions for the structure.}.

\begin{figure}
  \centering
  \igr[width=.45\textwidth,clip]{fig5}
  \caption{(color online) Force as function of average bond length, $T=\correlator{\theta_j}$.}
  \label{fig:forces}
\end{figure}

To shed light on the effect of straining the chains, we next investigate the energies
that are related to different types of movement by analyzing the eigenmodes and
eigenvalues of selected blocks of $\W$. Especially, we can consider the local motion of
individual atoms or groups of atoms, freezing all other degrees of freedom, by picking
the corresponding parts of $\W$. For a single atom this amounts to the on-site $3\times
3$ blocks. The square root of the positive eigenvalues of the reduced matrix, which we
call local energies, give the approximate energy of a solution to the full $\W$ that has
a large overlap with the corresponding eigenmode, if the coupling to the rest of the
dynamical matrix is low. The negative eigenvalues of a block are ignored since they
correspond to motion that is only stabilized by degrees of freedom outside the block.

\begin{figure}
  \centering
  \includegraphics[width=.45\textwidth,clip]{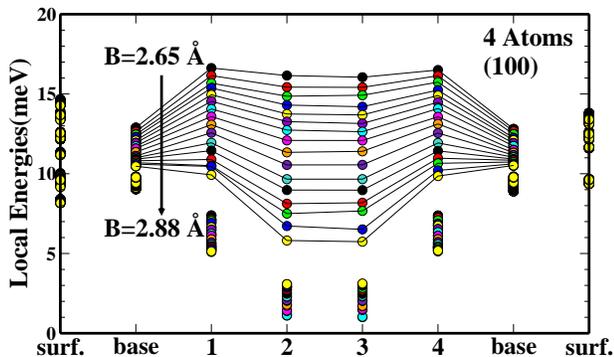}
  \caption{(color online) Local energies of a 4 atom chain between two (100)-surfaces
  at different strains. The largest
  eigenvalues are connected by a line to guide the eye.}
  \label{fig:onsite}
\end{figure}

The behavior of the dynamical matrix in terms of local energies, is relatively
straightforward, as illustrated in \figref{fig:onsite}. When the bonds are strained they
are also weakened. In the Central Chain the local energies are quickly reduced with
increased strain ($\approx 65\%$ decrease) while the dynamical matrix of the surfaces is
hardly affected. The Base and the first atom of the Chain fall in between these two
extremes with a $20\%$ and $40\%$ decrease, respectively. The middle bonds in the
Central Chain are the ones that are strained and weakened the most when the surfaces are
moved apart. It is also where the chain is expected to break\cite{Velez08}. Most
interestingly, we note that at least one jump in the on-site local energies occur when
moving from the Surface to the Central Chain.

\begin{figure}
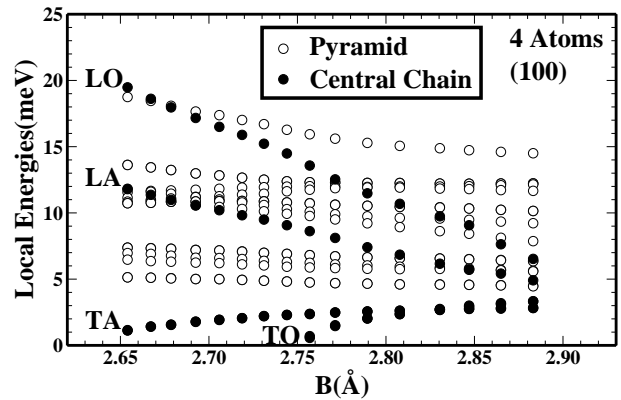

  \centering
  \igr[width=.45\textwidth,clip]{fig8}
  \caption{Local energies for selected blocks of $\W$ (Pyramid/Central Chain) plotted vs.
  the average bond length in the 4 atom chain. Since the central chain in this case consists
  of 2 atoms we can classify the eigenvectors as LO: longitudinal optical, LA: longitudinal
  acoustic, TO: transverse optical (doubly degenerate) or TA: transverse acoustic (doubly degenerate).}
  \label{fig:chainpyramid}
\end{figure}

In \figref{fig:chainpyramid} we see how motion parallel to the chain is at higher
energies than perpendicular motion, and that the LO type motion of the ABL modes has the
highest energy. We also see that the local energies of the ABL/LO type motion moves past
the local energies of the Pyramid as the strain is increased. In this way the ABL/LO
modes can in some sense act as a probe of the contacts.

\subsection{Mode life-times and $Q$-factors}
\label{sec:localisation}

\begin{figure}[h]
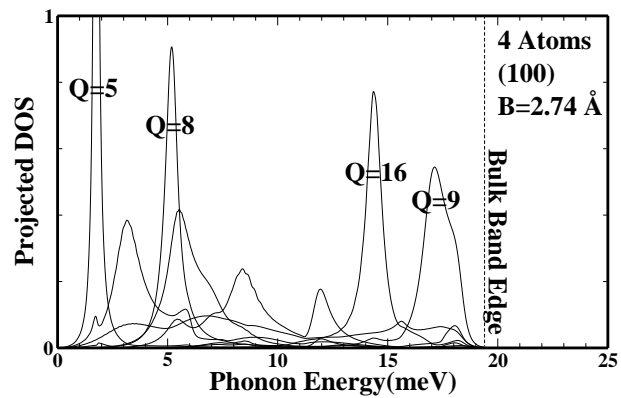

  \centering
  \igr[width=.45\textwidth,clip]{fig9}
  \caption{Projected DOS onto a representative selection of the vibrational
  modes of the device region that has a large overlap with the added
  (Chain+Base) region.}
  \label{fig:all_modes}
\end{figure}

We next investigate the modes of the finite chain systems. An example is given in
\figref{fig:all_modes} which depicts the projected DOS for a chain with 4 atoms at an
intermediate strain. Notice the large variation in the width of the peaks. The peaks
with a low width correspond to modes that have the largest amplitude in the Chain, while
the peaks with a large width correspond to modes with large amplitude on the Base and
Surface. Since this type of system has no natural boundary between 'device' and 'leads'
we will have large variation in the harmonic damping no matter where we define such a boundary.

\begin{figure}
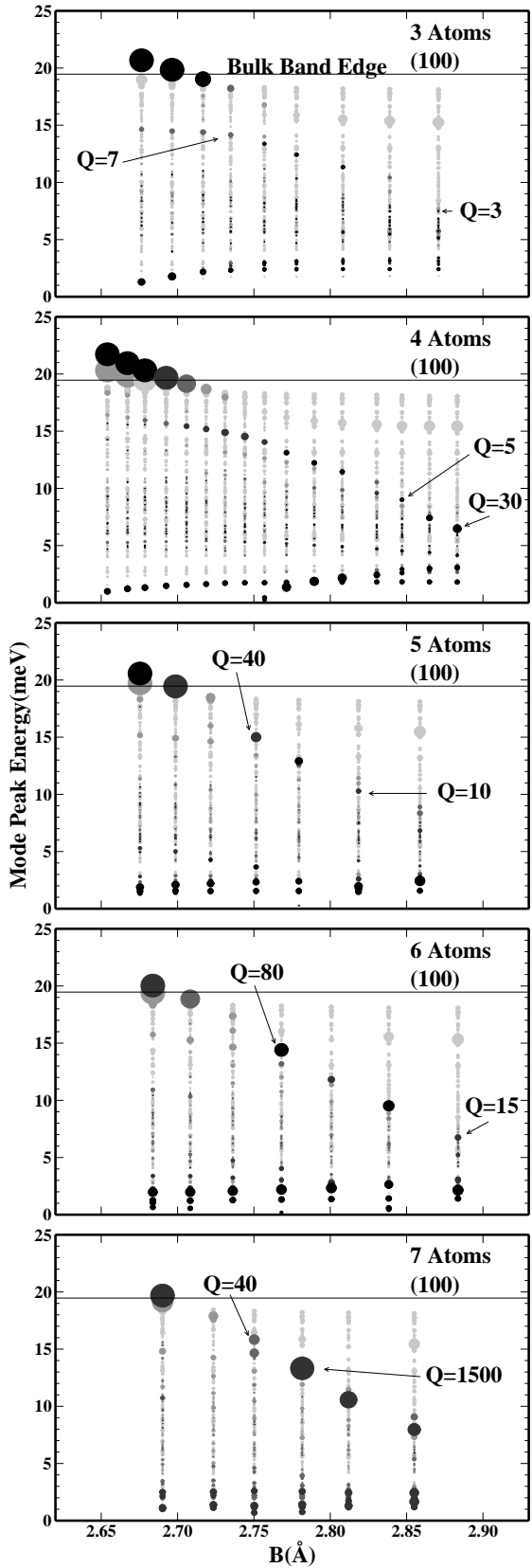

  \centering
  \includegraphics[width=.41\textwidth,clip,trim=-17 0 0 0]{fig10a}
  \includegraphics[width=.41\textwidth,clip,trim=-17 0 0 0]{fig10b}
  \includegraphics[width=.41\textwidth,clip,trim=0 0 0 0]{fig10c}
  \includegraphics[width=.41\textwidth,clip,trim=-17 0 0 0]{fig10d}
  \includegraphics[width=.41\textwidth,clip,trim=-17 0 0 0]{fig10e}
   \caption{The vibrational modes for chains with 3-7 atoms between two 100-surfaces.
   The center of the disks are positioned at the peak of the projection of
   vibrational DOS on the mode in question. The area of a disk is proportional
   to the $Q_\lambda$, but is limited to what corresponds to a $Q$-factor of 250.
   he gray level, that ranges from light gray to black in 4 steps signifies
   that $s_\lambda\in [0,2[\text{(light gray)},[2,4[,[4,6[,[6,8[$ or $[8,\infty[$(black).}
\label{fig:3_7_modes_overview}
\end{figure}

In \figref{fig:3_7_modes_overview} we present the $Q$-factor, spatial localization and
peak energy of all modes for chains with 3-7 atoms between (100) surfaces. These are the
main result of this article. \Tabref{tab:variation} shows the same information in an
alternative form.  We now proceed to an analysis of these results.

The ABL modes are of special interest. These modes have been identified by previous
theoretical and experimental studies as the primary scatterers of
electrons\cite{Agrait2002a,Frederiksen2004,Frederiksen2007,Viljas2005,Paulsson2005,Hobi2008,Hihath2008}.
The ABL modes are easily identified in \figref{fig:3_7_modes_overview} since they have
the highest energy of the modes that are spatially localized to the central chain (black
or dark gray on the figure). Modes corresponding to transverse motion of the central
chain are also clearly visible. These modes are energetically and spatially localized,
but are of limited interest because of a low electron-phonon coupling.

Certain ABL modes are very long-lived. At low strains, ABL modes lie
outside the bulk (and surface) band and have, in our harmonic
approximation, an infinite $Q$-factor. In reality the $Q$-factor will be limited by electron-phonon
and anharmonic interactions. At higher strain the ABL
modes move inside the bulk band and one observes a great variation
in the corresponding $Q$-factors. When the peak energy lies inside the bulk band there exists modes in the bulk with the same energy and it will mostly be the structure of the connection between the bulk crystal and the chain that determines the width of the peak.

The long chains tend to have longer lived ABL/LO modes due to the larger ratio between the
size of the Central Chain and the size of its boundary. The 7-atom chain is especially
interesting since it has an ABL/LO type mode with a damping of $5~\mev$ at one strain, while
at another strain the ABL/LO mode has a damping of  300~\mev, i.e., more than one
order-of-magnitude variation in the harmonic damping of the primary scatterer of electrons due
to only a 0.03~$\Ang$ change in the average bond length!

The largest damping of an ABL-mode for these systems is $\del\approx 1~\mev$, which is
still significantly lower than the $\approx 20~\mev$ band width. This can be attributed
to fact, noted above, that there always exists a large mismatch in local energies moving
from the central part of the chain to the rest of the system (see \figref{fig:onsite}).

Previous studies by Frederiksen~\etal\cite{Frederiksen2007a} obtained a rough estimate
for the variation of the non-electronic(harmonic and anharmonic) damping of 5-50~$\muev$ for the longer chains by fitting
the experimental IETS signals of Agra\"\i t~\etal\cite{Agrait2002} to a model calculation. The estimated peak energies lie well within the bulk band for all the recorded
signals. The reason the non-electronic damping rate can be extracted is because the excitation
of vibrations and damping of vibrations through electron-hole creation are both
proportional to the strength of the electron-phonon coupling. This means that the step
in the experimental conductance, when the bias reaches the phonon energy, can be used to
estimate strength of the electron-phonon interaction and thereby the electron-hole pair
damping. The slope in the conductance beyond this step can then be used to extract the total
damping. By subtracting the electron-hole pair damping from the total damping we get an estimate of the sum of the sum of harmonic and anharmonic contributions to the damping.

The estimate in Ref. [\onlinecite{Frederiksen2007a}] agrees well with our lowest damping of
5~\muev. The highest damping we have found was $\approx 400~\muev$ found for the 6 atom
chain which is an order of magnitude larger than the upper limit of Ref.
[\onlinecite{Frederiksen2007a}].  We believe that this discrepancy can be largely
attributed to the difficulty in extracting the necessary parameters from experiments
when the harmonic damping is large. Furthermore, for the 6-7 atom chains we observe that
the high damping occurs at low strain, where the electron-phonon coupling is
weak\cite{Agrait2002}.

\begin{figure}[h]
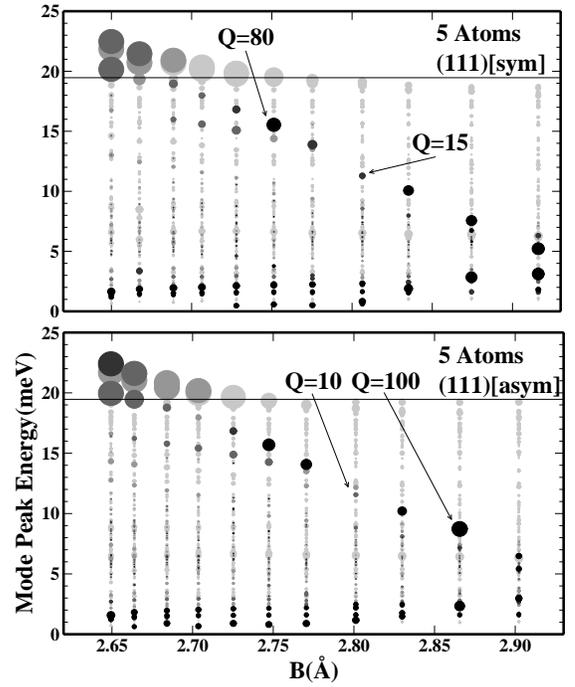

  \centering
    \includegraphics[width=.41\textwidth,clip,trim=-17 0 -2 0]{fig11a}\\
    \includegraphics[width=.41\textwidth,clip]{fig11b}
\caption{The vibrational modes for 5 atom chains between two (111) surfaces. [Top] Symmetric pyramids. [Bottom] Asymmetric pyramids (one atom added to one of
the pyramids). The area of a disk is proportional to the $Q_\lambda$, but is
limited to what corresponds to a $Q$-factor of 250. The gray level, that ranges
from light gray to black in 4 steps signifies that $s_\lambda\in [0,2[\text{(light gray)},[2,4[,[4,6[,[6,8[$ or $[8,\infty[$(black).}
  \label{fig:5_modes_overview}
\end{figure}

\begin{table}[h]
  \centering
  \begin{tabular}{| c | c | c | c |}
    \hline
    Chain & $Q_\lambda$ & $\del$(\muev) & $\tau_\lambda$(\ps)\\
    \hline
    3(100) & 3-7 & 800-1200 & 0.5-0.8 \\
    \hline
    4(100) & 5-30 & 100-900& 0.7-7\\
    \hline
    5(100) & 10-40 & 200-500& 1.3-3 \\
    \hline
    6(100) & 15-80 & 90-400 & 1.6-7\\
    \hline
    7(100) & 40-1500 & 5-300 & 2-130\\
    \hline
    5(111)(symmetric) & 15-80 & 40-400 & 1.6-16\\
    \hline
    5(111)(asymmetric) & 10-100 & 40-800 & 0.8-16\\
    \hline
  \end{tabular}
  \caption{The variation of the $Q_\lambda$, $\del$ and $\tau_\lambda$ of the ABL/LO-modes.  For Chains with 3-7 between (100) surfaces and for Chains with 5 atoms between (111) surfaces but with slightly different Bases. The strains where the peak energy of the ABL/LO-mode falls close to or outside the bulk band edge have been disregarded. }
  \label{tab:variation}
\end{table}
There are two main differences between the (100) and the (111) systems. The first
difference is that the (111)-systems have ABL/LO-modes that are more long-lived compared to
the (100)-systems (see \tabref{tab:variation} and \figref{fig:5_modes_overview}). The
second difference is the behavior of the localized modes close to the band edge (See
\figref{fig:5_modes_overview}). The modes with energies outside the bulk band in the
(111) systems are less spatially localized compared to the (100) case. At low strain,
the (111)-chain have ABL/LO-modes extending further into the Base and Surface than the
(100)-chain.

There are certain general features of how the damping evolves with
strain that are easily understood. Modes with peak energies in the
range $16-19~\mev$ in generel have a very high damping while those in
the range $14-16~\mev$ have very low damping. This correlates well
with the bulk DOS for gold (see e.g. Ref. [\onlinecite{Treglia1985}]). The optical peak in the bulk DOS
corresponds to strong damping while the gap between optical and
acoustical modes correspond the the range of low damping.

To sum up, localized modes occur at low strain where the bonds in the chain are very
strong, and give rise to frequencies close to or outside the bulk band edge. Inside the
bulk band strong localization is still possible for the long chains, especially the
7-atom chain. This requires, however, that the coupling between the Central Chain and
the surface is weak at the typical frequency of the ABL/LO mode due to the structure of the
connection. The behavior depends strongly on the detailed structure of the base and the state of strain, but some general features can be related to the the bulk DOS.

\section{Conclusion and Discussion}
\label{sec:conclusions}

We have presented a study of the harmonic damping of vibrational modes in gold
chains using a method that uses ab-initio parameters for the chains and empirical
parameters for the leads. We have focused on the ABL/LO modes that interact strongly
with electrons and are thereby experimentally accessible through $IV$ spectroscopy. We
provide an estimates for the damping of ABL/LO-modes from ab-initio
calculations as a function of strain for a wide range of gold chain systems. The
calculations of the ABL-phonon damping rates agree well with earlier estimates, found by
fitting a model to experimental inelastic signals\cite{Frederiksen2007,Agrait2002}.

We have found the that the values of the harmonic damping for the ABL
modes can vary by over an order of magnitude with strain. Even with
small variations in the strain, the harmonic damping can exhibit this
strong variation. This extreme sensitivity may
explain the large variations seen experimentally in different
chains.

The range of the harmonic damping also depends strongly on the number of atoms in the chain since
we see a clear increase in localization going from a 6- to a 7-atom chain. The chain
with 7 atoms really stands out, since it, in addition to having very localized modes in generel, it also has the greatest variation in harmonic damping. This strong variation in
the harmonic damping of the ABL/LO-modes, that depends on the details of the structure, suggest
that accurate atomistic calculations of the vibrational structure is necessary to
predict the inelastic signal.

All types of chains were found to have ABL/LO-modes tha lie outside the bulk phonon band at
low strain. These modes are expected to have very long life-times since the harmonic
damping is zero. Signatures of the rather abrupt change in the damping of the ABL/LO-modes
when strained have not been discussed in experimental literature so far. We believe this is due to the common experimental techniques for producing these chains heavily favor strained chains. The ABL/LO-mode
life-time may be set by the coupling to the electronic system (electron-hole pair
damping). Indeed, even inside the bulk band the electron-hole pair damping can be of the
same order as the harmonic damping. For example, a $\gamma_{\lambda~eh}\approx
50-80~\muev$ was found for a 4-atom\cite{Paulsson2005} and a
7-atom\cite{Frederiksen2007a} chain, which we can compare with $100-900~\muev$ and
$5-300\muev$ found above for the harmonic vibrational damping. Thus the damping can in
certain cases be dominated by the electron-hole pair damping for frequencies even inside the
bulk band.

Finally we find a difference in the the damping of ABL/LO modes in chains between (100)-
and (111)-surfaces. For the investigated 5-atom chains there is both a marked difference
in the strength of damping and in the variation of the damping with strain. It might be
possible to distinguish between (100) and (111) pyramids experimentally due to this
difference. The ABL/LO modes will have strong coupling to the bulk at certain energies,
characteristic of the pyramid type. This in turn, results in broadening/splitting of the
modes depending on whether the characteristic energies are inside or outside the bulk
band. This broadening/splitting would be detectable in the $IV$-curve since it is related to
the characteristics of the conductance step at the peak energy of the vibrational mode.
Finding $IV$-curves at different strains could thereby serve as a fingerprint of the
specific way the chain is connected to the surroundings. Hihath \etal~have demonstrated
that such measurements are indeed possible on a single-molecule
contact\cite{Hihath2008}.

The techniques used in this paper can be combined with electronic transport calculations
to predict the inelastic signal in the $IV$ characteristic of a system. This will be
done in future work, where we will also eliminate the use of the empirical model for the leads and use
ab-initio parameters for the entire system.

\section{Acknowledgements}

The authors would like to thank Thomas Frederiksen for helpful discussions and
Nicolas Agra\"it showing his unpublished experimental results.
A. P. Jauho is grateful to the the FiDiPro program of the Finnish Academy. Computational resources were provided by the Danish Center for Scientific Computing (DCSC).

\appendix

\section{Constructing the Dynamical Matrix}
\label{sec:constr-dynam-matr} In this subsection the details of how we constructed the
dynamical matrix are presented. The dynamical matrix must be symmetric and obey momentum
conservation. Momentum conservation, in this context, means that when an atom is
displaced the force on the displaced atom equals minus the total force on all other
atoms. We ensure momentum conservation by setting the on-site $3\times 3$ matrix to
minus the sum of the force constant coupling matrices to all the other atoms. This
method for regularizing the dynamical matrix was previously used by
Frederiksen~\etal\cite{Frederiksen2007a}, and generally improves on the errors introduced
in the total energy when displacing atoms relative to the underlying computational grid
(the DFT egg-box effect). We calculate off-diagonal coupling part of the force constant
matrix was calculated with a finite difference scheme using a displacement, $Z$, of
$0.02~\Ang$ in the $x$,$y$ and $z$ directions for all atoms in the Chain and Base.

To improve the accuracy further, the forces were calculated for both positive and
negative displacement. If $i$ and $j$ are degrees of freedom situated inside the DFT
region we therefore perform 4 independent calculations of $\W_{ij}=\W_{ji}$, since $\W$
is a symmetric matrix. In the end we use the average of the force constant from these 4
calculations
\begin{equation*}
  \W_{ij}=\frac{\hbar^2}{\sqrt{m_im_j}}
  (\frac{F_{ij+}}{Z}+\frac{F_{ji+}}{Z}-\frac{F_{ij-}}{Z}-\frac{F_{ji-}}{Z})/4\,,
\end{equation*}
where e.g. $F_{ij+}$ denotes the force on $i$ due to a positive displacement of $j$. If
$i$ is inside the DFT region and $j$ is not, the coupling is calculated as an average of
2 force constants
\begin{equation*}
  \W_{ij}=\frac{\hbar^2}{\sqrt{m_im_j}}(\frac{F_{ji+}}{Z}-\frac{F_{ji-}}{Z})/2\,.
\end{equation*}

If an atom was close to a periodic image of another atom (less than half the unit cell
length in any direction) the force between these atoms was set to zero to avoid
artifacts of the periodic calculational setup. The empirical model was used to calculate the coupling between the surface atoms. After all coupling elements were found the on-site elements were calculated for the system as a whole.

\section{Convergence}
\label{sec:convergence}
\subsection{Convergence Parameters}
\label{sec:conv-param} In the calculations there are several convergence parameters, and here we provide an
overview.

There are three important length-scales in the calculations: $L_1$,$L_2$ and $L_3$. We assume that
when two atoms are further apart than $L_1$, the coupling elements between them
vanishes. $L_2$ is the correlation length for properties that do not have an
energy-dependence, like forces, equilibrium positions and total energies, while $L_3$ is
the assumed correlation length for properties that do have an energy-dependence, like
the surface Green's function, vibrational DOS etc. $L_3$ always needs to be larger than
$L_2$, $L_3>L_2$, but the specific size needed depends on the required energy resolution. $L_2$
determines the $k$-point sampling used in the DFT-calculations and $L_3$ the $k$-point
sampling used in the calculation of the surface Green's function. In each case the
number of $k$-points used one direction is chosen to be the smallest integer, $i$, such that $i>\frac{L}{a})$ where $a$ is the size of
the calculational cell in that direction. The DFT $k$-point sampling used is dense
enough to ensure that $L_2>23~\Ang$ for all calculations.

In the calculation of the Green's functions we introduced a finite artificial
broadening. This broadening, $\eta$, was divided into a small broadening of the device
region, $\eta_C$, and a large broadening for the leads, $\eta_L$. The reasoning behind
this is that the density of states is much more smooth in the bulk-like regions far away
from the chain. A large $\eta_L$ has the advantage that it reduces the need for
$k$-point sampling drastically. Without a small $\eta_C$ we would not be able to discover very
sharp peaks in the DOS. To reliably find the modes of the system it is also important
that the energy spacing, $\Delta E$ is on the same level or smaller than $\eta_C$.

The artificial broadening limits how large life-times we can resolve. This is why we in
the following write the upper limit to the life-time introduced by the artificial
broadening.

A final convergence parameter is the position of the interface between DFT and empirical
model parameters for the dynamical matrix. This is a very important parameter since the error
introduced by having this interface relatively close to the chain is what limits the precision of the calculations.

\subsection{Test of Convergence}
\label{sec:test-convergence}

\begin{figure}[ht]
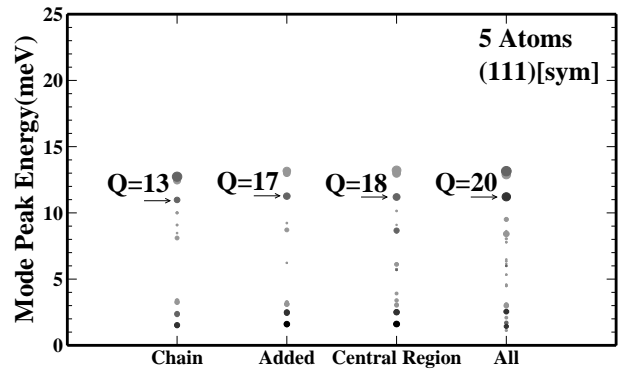

  \centering
  \igr[width=.45\textwidth,clip]{fig12}
  \caption{Modes of the device region with a different DFT/EM interface. The label designate the region treated with DFT, where the 'Added' region is the one used in the main part of the calculations and 'All' is fully ab-initio. See \figref{fig:5_modes_overview} to see what the color and size signify. On this plot modes with $s_\lambda\in[0,2[$ are suppressed.}
\label{fig:convergence2}
\end{figure}

Next we present the tests that have been carried out to ensure that the calculations in this
article are sufficiently converged. The convergence for the SIESTA basis set and the
size of the finite displacement used in the finite difference calculations was already
tested for the same type of systems by Frederiksen~\etal\cite{Frederiksen2007a}.

So here we first examine the convergence of the DFT calculations of the dynamical
matrix. A calculation for a 4 atom chain between (100)-surface was done with improved
values for the important DFT convergence parameters. The mesh cutoff was increased from
150 to 200~$\Ry$, and the $k$-point sampling was increased from $2\times 3$ to
3$\times$4. For this change in parameter we obtained a maximal difference of $0.2~\mev$,
when comparing the square root of the sorted array of eigenvalues of the dynamical
matrix. This is a negligible size since the average value of the eigenvalues is about
10~\mev. The $k$-point sampling in the DFT calculations proved crucial for the structure
of the strained systems, since gamma-point calculations resulted in different structures
(different bonds weakened at high strain) with very large life-times.

The perturbation length used in our calculations was $L_1=4.08\Ang$, which is the same
as next-nearest neighbor distance. The magnitude of any next-nearest-neighbor coupling
matrix, defined as $|\X|=\sqrt{\sum_{ij} X_{ij}^2}$ was never larger than $15\%$
compared to the magnitude of any nearest-neighbors coupling matrix. The error introduced
by this truncation is smaller than the one introduced by using the empirical model for
the dynamical matrix.

For the calculation of the DOS we gradually improved $L_3$, $\eta_L$ and $\eta_C$ and
found that the DOS was converged using $\Delta E=10~\muev$, $\eta_L=100 ~\muev$($7$~ps),
$\eta_C=10 ~\muev$($70$~ps) and $L_2=200 ~\Ang$(68$\times$68 $k$-points) except in one
calculation for the 7 atom chain we needed the life-time of one very sharp peak. This
required a better resolution using $\Delta E=1~\muev$, $\eta_L=10~\muev$($70$ ps),
$\eta_C=1~\muev$($700$~ps) and $L_2=400~\Ang$ (136$\times$136 $k$-points).

Finally, we have considered how much the interface between the DFT parameters and the
empirical parameters affect our results. In \figref{fig:convergence2} we show a study
where we vary the position of this interface. We find that our calculation of the
$Q$-factor and the spacial localization is converged to about the first significant
digit for modes that are spatially localized to the Central Chain. We judge that this
is what mainly sets the limit of accuracy of in our calculations.

\section{Definition of the Modes for an Open System}
\label{sec:finding-modes-peak}
The starting point is the modes of a closed system, namely, the eigenmodes of $\W$. The most important requirement, for the defition of modes in the case of the open system, is that these modes become the modes of the isolated system in the limit of zero coupling between
the device region and the leads.

The following definition fulfills this condition. A 'mode' is defined as a (complex)
eigenvector  $u_\lambda$ of $\M_{DD}(\epsilon^*)$  (and $\D_{DD}(\epsilon^*)$) that
fulfills,
\begin{equation}
\label{eq:condition1}
\real\{u_\lambda^\dagger\D_{DD}(\epsilon^*)u_{\lambda}\}=0\,,
\end{equation}
and
\begin{equation}
\label{eq:condition2}
\frac{\df}{\df \epsilon} \real\{u_\lambda^\dagger\D_{DD}(\epsilon)u_{\lambda}\}|_{\epsilon=\epsilon^*}>0
\end{equation}
for some energy, $\epsilon^*$, corresponding to a peak in DOS.

\begin{figure}[t]
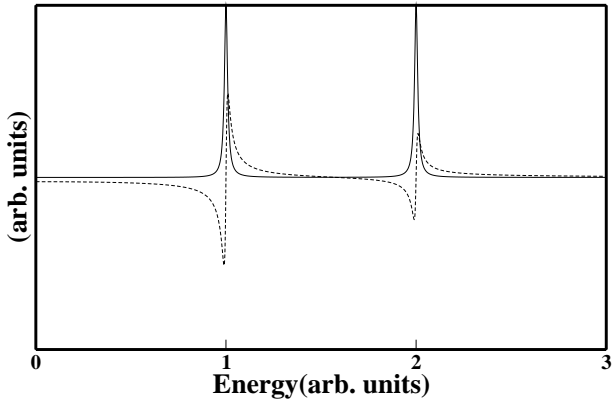

  \centering
  \igr[width=.45\textwidth,viewport=0 0 355 250,clip]{fig13}
  \caption{Example $\real\D$ and $n(\epsilon)$, dashed and solid line respectively, for a Green's function with two poles at 1 and 2 with a 0.1 broadening. We see that the values where the real part is zero only correspond to peaks in the density if the slope positive.}
  \label{fig:two_modes}
\end{figure}

An illustration of these two conditions is given in \figref{fig:two_modes}. In practice,
the modes are found from the number of positive eigenvalues of $\D_{DD}$ evaluated at
each point of our energy-grid. If this number increases between two successive energies, $\epsilon$ and $\epsilon+\Delta \epsilon$,
the eigenmodes at these two energies are matched up. The eigenmode corresponding to the
eigenvalue that changes sign is then identified as a mode of the open system.

\bibliography{intro1,intro,goldwires,thermal,inelastic}

\end{document}